\documentclass[aps, preprint, prc, amsmath, amssymb, nofootinbib, superscriptaddress]{revtex4}

\usepackage[usenames]{color}
\usepackage[dvipsnames]{xcolor}
\usepackage{graphicx}
\usepackage{amsmath}
\usepackage{amsfonts}
\usepackage{amssymb}
\usepackage{mathrsfs}
\usepackage{bm}
\usepackage{verbatim}
\usepackage{cancel}
\usepackage{comment}
\usepackage[normalem]{ulem}
\usepackage{CJK}
\usepackage{float}
\usepackage{textcomp}
\usepackage{booktabs}

\usepackage{slashed}

\usepackage{hyperref}
\hypersetup{
  colorlinks=true,        
   linkcolor=ForestGreen,    
}

\usepackage{multirow,makecell}

\newcommand{\mhi}{M_{\text{hi}}}

\newcommand{\chn}[3]{{{}^{#1}\!{#2}_{#3}}}
\newcommand{\cs}[2]{\chn{#1}{S}{#2}}
\newcommand{\cd}[2]{\chn{#1}{D}{#2}}

\newcommand{\csd}{{\cs{3}{1}-\cd{3}{1}}}

\newcommand{\NNLO}{N$^2$LO}
\newcommand{\NNNLO}{N$^3$LO}

\graphicspath{{./figs/}}
\begin{document}

\begin{CJK*}{UTF8}{gbsn}
 
\title{Renormalization of proton-proton fusion in chiral effective field theory}

\author{Tai-Xing Liu (刘太兴)}
\affiliation{College of Physics, Sichuan University, Chengdu, Sichuan 610065, China}

\author{Rui Peng (彭锐)}
\affiliation{College of Physics, Sichuan University, Chengdu, Sichuan 610065, China}

\author{Songlin Lyu (吕松林)}
\email{songlin@scu.edu.cn}
\affiliation{College of Physics, Sichuan University, Chengdu, Sichuan 610065, China}

\author{Bingwei Long (龙炳蔚)}
\affiliation{College of Physics, Sichuan University, Chengdu, Sichuan 610065, China}

\date{October 13, 2022}

\begin{abstract}
Renormalization of proton-proton fusion is studied in the framework of chiral effective field theory. Strict perturbative treatment of subleading corrections is applied in the analysis. Possible enhancement of two-nucleon contact axial current operators is the focus of the study. We find evidence that supports a previous proposal in the literature to promote one of the contact axial current operators.
\end{abstract}

\maketitle

\end{CJK*}

\section{Introduction}

As the first reaction of the proton-proton ($pp$) chain, $pp$ fusion is the predominant process in conversion of hydrogen to helium in light stars like the Sun. Its rate is an essential ingredient in understanding stellar nucleosynthesis. However, the reaction cross section is difficult to measure in terrestrial laboratories; therefore, reliable theoretical prediction of it is often needed as one of the inputs for stellar models. In the present paper, we examine power counting of weak currents involved in this process  in chiral effective field theory (ChEFT), with renormalization-group (RG) invariance as the guideline.

At the hadronic level, the total cross section of $pp$ fusion consists of two essential elements: nuclear wave functions and axial current operators. Near the threshold, it can be schematically written as
\begin{equation}
\sigma\left(E\right) \propto \sum_{M} 
\lvert \langle \psi_d^M\rvert \vec{A}_- \lvert\psi_{pp} \rangle\rvert^2,
\label{eqn:approx_cross_section}
\end{equation}
where $\vec{A}_-$ denotes the axial current, $\psi_d^M$ ($\psi_{pp}$) the deuteron bound state ($pp$ scattering state), $M$ the $z$component of the deuteron spin, and $E$ the center-of-mass (CoM) energy. 

In the early investigations, both strong and weak interactions were phenomenologically constructed~\cite{Bethe:1938yy, Salpeter:1952ffc, Bahcall:1968wz, Kamionkowski:1993fr, Schiavilla:1998je}. As ChEFT developed, interests in $pp$ fusion were revived due to the prospect of quantifying its theoretical uncertainty in an EFT framework~\cite{Park:1998wq, Park:2002yp}. In the so-called hybrid approach, current operators were derived from ChEFT and various potential models were used to construct the nuclear wave functions. 

Full EFT calculations were actually carried out at first in pionless EFT~\cite{Kong:2000px, Butler:2001jj, Ando:2008va, Chen:2012hm, Behzadmoghaddam:2020pqr}. At next-to-leading order (NLO), however, a low-energy constant (LEC) is needed: $L_{1,A}$, which parametrizes the two-body axial current. Several means to determine $L_{1,A}$ were proposed in Refs.~\cite{Butler:2002cw, Savage:2016kon, De-Leon:2016wyu}.

Applications of ChEFT to both potentials and currents were performed in Refs.~\cite{Marcucci:2013tda, Acharya:2016kfl}. In these ChEFT calculations, power counting of potentials and currents are based on naive dimensional analysis (NDA). NDA has been shown to be inconsistent with RG invariance and various power counting schemes of nuclear forces have been proposed to meet the requirement of RG invariance~\cite{Birse:2005um, Birse:2007sx, Birse:2009my, Valderrama:2009ei, PavonValderrama:2011fcz, Long:2011qx, Long:2011xw, Long:2012ve, PavonValderrama:2019lsu, vanKolck:2020llt, Zhou:2022loi}. RG analysis of the nuclear currents in ChEFT was pioneered by Ref.~\cite{PavonValderrama:2014zeq}, based on the short-range behavior of two-nucleon wave functions. The strategy of using RG for power-counting nuclear currents was also applied in studying beyond-standard model physics in nuclei~\cite{Cirigliano:2018hja, Oosterhof:2019dlo, Yao:2020olm}. For different points of view towards RG invariance in the context of chiral nuclear forces, we refer to 
Refs.~\cite{Epelbaum:2009sd, Epelbaum:2006pt, Epelbaum:2018zli, Gasparyan:2021edy}. 

We examine power counting of axial current operators especially for the process of $pp$ fusion. Since $\Delta$-less potentials are used in the present work, we set the breakdown scale $\mhi \simeq \delta \simeq 300$ MeV--- the $\Delta$-nucleon mass splitting--- to estimate uncertainties. Besides using RG invariance as a guideline, we treat higher-order potentials in perturbation theory in the same manner as they were studied in Refs.~\cite{Valderrama:2009ei,PavonValderrama:2011fcz,Long:2011qx, Long:2011xw, Long:2007vp, Long:2012ve, SanchezSanchez:2017tws, Wu:2018lai, Peng:2020nyz, Peng:2021pvo}, as opposed to lumping them altogether with the leading-order (LO) potential in the Schr\"odinger equation~\cite{vanKolck:2020llt}. 

The paper is organized as follows. In Sec.~\ref{sec:ppscat}, we demonstrate how to deal with the $pp$ interaction by calculating the $pp$ $\cs{1}{0}$ phase shifts up to next-to-next-to-leading order (N$^2$LO). We then discuss the nuclear matrix element of $pp$ fusion in Sec.~\ref{sec:rme}, including relevant axial current operators and the deuteron wave function. This is followed by results and discussions in Sec.~\ref{sec:results}. Finally, a summary is offered in Sec.~\ref{smry}.

\section{Proton-proton scattering\label{sec:ppscat}}
We describe near-threshold $pp$ scattering where the energy is so low that the Coulomb potential must be fully iterated. For discussions on perturbative treatment of the Coulomb potential in the context of pionless and cluster EFTs, we refer to Refs.~\cite{Konig:2015aka, Kirscher:2015zoa, Konig:2016iny}.

The full $T$ matrix in the presence of the strong and Coulomb interactions can be divided into two parts: the pure Coulomb part $T_{c}$ and the modified strong amplitude $\widetilde{T}_{sc}$~\cite{Goldberger:1964ny}. We begin by introducing the Coulomb propagator:
\begin{equation}
    G_c^{\pm}(E) = \frac{1}{E - H_0 - V_c \pm i\epsilon},
\end{equation}
where $H_0$ is the free Hamiltonian, $V_c$ is the Coulomb potential, and the CoM energy $E = p^2/m_N$ with the nucleon mass $m_N = 939$ MeV. The Coulomb amplitude $T_c(\vec{p}, \vec{p}\,')$ is defined as~\cite{Kong:1999sf}
\begin{equation}
T_{c}(\vec{p}\,', \vec{p}) = \langle\vec{p}\,'\left\vert V_c \right\vert \psi_c^+(\vec{p})\rangle \, .
\end{equation} 
Here, the incoming ($\psi_c^{-}$) and outgoing ($\psi_c^{+}$) 
Coulomb wave functions are given by
\begin{equation}
    \vert \psi_c^{\pm}(\vec{p})\rangle = \left(1 + G_c^{\pm} V_c \right)\vert \vec{p}\rangle \, .
\end{equation}
Operator $T_{sc}$ is defined by iterating the strong potential $V_\text{str}$ through $G_c(E)$:
\begin{equation}
T_{sc} = V_\text{str} + V_\text{str} G_c(E) T_{sc} \, ,
\label{eqn:TscDef}
\end{equation}
and $\widetilde{T}_{sc}(\vec{p}\,', \vec{p})$ is the matrix element of $T_{sc}$ between $\psi_c^+$ and $\psi_c^-$:
\begin{equation}
\widetilde{T}_{sc}(\vec{p}\,', \vec{p}) \equiv \langle \psi_c^-(\vec{p}\,')\left\vert T_{sc} \right\vert  \psi_c^+(\vec{p})\rangle \, .
\label{Tsc:expression}
\end{equation}

$T_c$ and $T_{sc}$ can be projected onto partial waves in a fashion similar to their strong-interaction counterparts. More specifically, $\widetilde{T}_{sc}(p, p)$ for $\cs{1}{0}$ is related to the strong phase shift $\delta_{sc}(p)$ as follows:
\begin{equation}
\widetilde{T}_{sc}(p, p) = -\frac{4\pi}{m_N}e^{2i\delta_c(p)}\frac{e^{2i\delta_{sc}(p)}-1}{2ip}\, ,
\end{equation}
where $p$ is the CoM momentum and $\delta_c(p)$ the Coulomb phase shift. We will restrict ourselves to the $^1S_0$ channel of $pp$ interaction because the $P$-wave contribution to near-threshold $pp$ fusion is smaller than the $S$-wave contribution by several orders of magnitude~\cite{Marcucci:2013tda,Acharya:2019zil}. We drop the subscript of orbital angular momentum to simplify the notation.

The technique presented in Refs.~\cite{Vincent:1974zz, Walzl:2000cx} is adopted to calculate the strong phase shift $\delta_{sc}(p)$. An artificial infrared cutoff in coordinate space $R_p$ is introduced, beyond which the strong potential is neglected. One expects $\delta_{sc}$ to be independent of $R_p$ as long as $R_p$ is much larger than the range of $V_\text{str}$. We have verified that, when $R_p$ is chosen to be $10$ fm, the relative errors of the phase shifts $\delta_{sc}(p)$ are smaller than $10^{-3}$. The $pp$ scattering wave function $\psi_{pp}(\vec{r}; \vec{p}\,)$ will be constructed by this method. It is useful to show the spin and isospin structure of $\psi_{pp}(\vec{r}; \vec{p}\,)$~\cite{Schiavilla:1998je}:
\begin{equation}
\psi_{pp}(\vec{r}; \vec{p}) = 4\pi\sqrt{2}e^{i\delta_{sc}}\frac{\chi_0(r;p)}{pr}Y^*_{00}(\hat{p})Y_{00}(\hat{r})\eta^0_0\zeta^1_1
\label{pp:function},
\end{equation} 
where $\chi_0(r;p)$ is the radial wave function and $\eta_S^{M_S}$ ($\zeta_{T}^{M_T}$) is the spin (isospin) piece of the wave function, with the $z$ component $M_S$ ($M_T$). 

The power counting for the neutron-proton ($np$) $\cs{1}{0}$ interaction explained in Ref.~\cite{Long:2012ve} is our starting point for the strong potentials. Later, other schemes were proposed to improve the convergence of ChEFT in $\cs{1}{0}$~\cite{Long:2013cya, SanchezSanchez:2017tws, Peng:2021pvo, Mishra:2021luw, Ren:2017yvw}, but they are aiming at momenta much higher than 
those we are concerned with in the present paper. Following Ref.~\cite{Long:2012ve}, we expand the $\cs{1}{0}$ potential $V_\text{str}$ up to N$^2$LO:
\begin{align}
V_\text{str}^{(0)}(p',p) &= V_{1\pi}(p\,',p) + C^{(0)},\\
V_\text{str}^{(1)}(p',p) &= C^{(1)} + \frac{1}{2}D^{(0)}(p'^2+p^2),\\
V_\text{str}^{(2)}(p',p) &= V_{2\pi}(p',p) + C^{(2)} + \frac{1}{2}D^{(1)}(p'^2+p^2) + \frac{1}{2}E^{(0)}p'^2p^2,
\end{align}
where the LECs $C$ and $D$ are formally expanded at each order: $C = C^{(0)} + C^{(1)} + C^{(2)}$ and $D = D^{(0)} + D^{(1)}$. To regularize the ultraviolet part of potentials, we use a separable Gaussian regulator: 
\begin{equation}
V^{\Lambda}(p',p) \equiv \exp\left(-\frac{p'^{\,4}}{\Lambda^4}\right)V(p',p)\exp\left(-\frac{p^{\,4}}{\Lambda^4}\right).
\end{equation}

Unlike in the $np$ sector, the $pp$ contact interactions are renormalized by the Coulomb force at short distances. On the other hand, the OPE potential--- the long-range part of the strong interactions--- is unchanged from $np$ to $pp$. Because OPE behaves similarly to the Coulomb force for $r \to 0$, where $r$ is the internucleon distance, one expects the addition of the Coulomb force only to change the renormalization of the contact terms modestly and the power counting for the $\cs{1}{0}$ $pp$ contact terms to remain in the same pattern as that for $np$. In addition to this argumentation, we check the power counting against RG invariance by verifying numerically that $\delta_{sc}$ is independent of the cutoff value at each order.

The perturbative treatment of higher-order potentials may be most conveniently explained by a generating function. We introduce an auxiliary parameter $x$ and define a potential in the form of $x$ polynomials, with $V_\text{str}^{(n)}$ as the coefficient of $x^n$:
\begin{equation}
 V_\text{str}(p',p;x) = V_\text{str}^{(0)}(p',p) + xV_\text{str}^{(1)}(p',p) + x^2V_\text{str}^{(2)}(p',p) + \mathcal{O}(x^3) \, .
\label{eqn:VstrExpan}
\end{equation}
This potential results in an $x$-dependent amplitude, $\widetilde{T}_{sc}(p', p; x)$, whose Taylor expansion in $x$ leads to the desired correction to the LO amplitude $\widetilde{T}_{sc}^{(0)}(p', p)$:
\begin{equation}
\widetilde{T}_{sc}(p',p;x) = \widetilde{T}_{sc}^{(0)}(p',p) + x\widetilde{T}_{sc}^{(1)}(p',p) + x^2\widetilde{T}_{sc}^{(2)}(p',p) + \cdots \, .
\label{eqn:TscExpan}
\end{equation}
One can follow the same suit to relate the EFT expansion of $\delta_{sc}$ to that of $\widetilde{T}_{sc}$.

In the numerical calculations, the following values are taken for various parameters: the fine-structure constant $\alpha$ = 1/137.036, the axial vector coupling constant $g_A$ = 1.29, the pion decay constant $f_{\pi}$ = 92.4 MeV, and the pion mass $m_{\pi}$ = 138 MeV. 

\begin{figure}[htbp]
    \centering
    \includegraphics[scale=0.8]{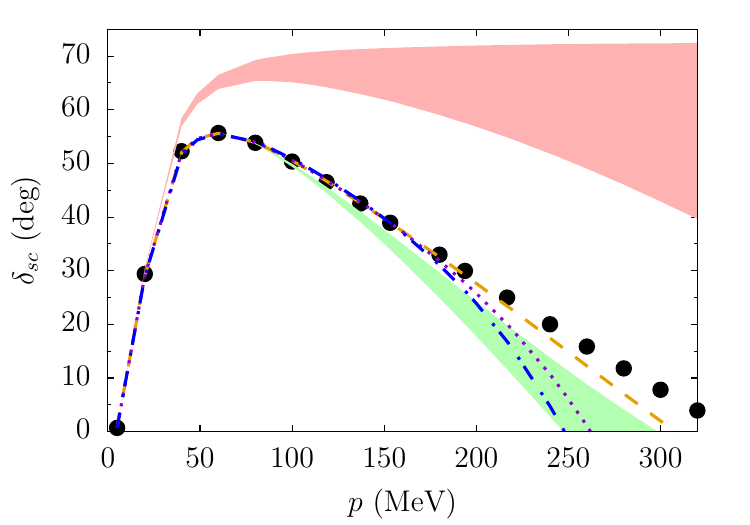}
    \caption{The $pp$ $^1S_0$ phase shift as a function of the CoM momentum $p$.
    The solid circles represent the empirical values from Ref.~\cite{NNonline}. The red and green bands represent the results at LO and NLO, respectively, from $\Lambda$ = 0.5 to 3.2 GeV. At {\NNLO}, $\Lambda$ = 0.5, 1.5, and 3.2 GeV are represented by, respectively, dashed, dotted, and dot-dashed curves.}
    \label{fig:pp1s0phase}
\end{figure}

To determine the LECs of $pp$ contact interactions, we fit $\tilde{T}_{sc}$ to the empirical values of $pp$ phase shifts provided by the partial-wave analysis in Ref.~\cite{NNonline}. At LO, the phase shift at CoM momentum $p = 5.0$ MeV is used as the input. At NLO and N$^2$LO, $p = 68.5$ and 153.2 MeV are added. The $^1S_0$ phase shifts up to N$^2$LO are shown in Fig.~\ref{fig:pp1s0phase}. The convergence of EFT expansion and the cutoff variation bands is similar to that for $np$ scattering presented in Ref.~\cite{Long:2012ve}. A shift from $\Lambda =$ 1.5 to 3.2 GeV smaller than that from 0.5 GeV to 1.5 GeV indicates the cutoff convergence for large $\Lambda$'s at N$^2$LO.
Near $p \simeq 200$ MeV, the {\NNLO} cutoff variation becomes comparable to that of the NLO band. This echos the slow convergence of $\cs{1}{0}$, as mentioned earlier.

\section{Axial currents and matrix elements\label{sec:rme}}

We at first use NDA to take stock of the axial current operators to be used in the paper. The weak current $\vec{A}$ for the two-nucleon system can be written in the plane-wave basis as:
\begin{equation}
    \langle \vec{P}\,'\; \vec{p}\,'\vert \vec{A}\vert\vec{P}\; \vec{p}\,\rangle=\vec{A}_{1B}(\vec{p}\,',\vec{p}; \vec{q}\,)(2\pi)^3\delta^{(3)}(\vec{p}\,' - \vec{p} - \frac{\vec{q}}{2})+\vec{A}_{2B}(\vec{p}\,',\vec{p};\vec{q}\,)
\end{equation}
where $\vec{p}$ ($\vec{p}\,'$) denotes the initial (final) relative momentum, $\vec{P}$ ($\vec{P}\,'$) the initial (final) total momentum, $\vec{q} = \vec{P}\,'-\vec{P}$ the momentum carried by the current, and $\vec{A}_{1B}$ ($\vec{A}_{2B}$) the one-body (two-body) current operators. Up to {\NNLO} in NDA, only one-body axial current operators contribute to the $pp$ fusion rate. When there is no ambiguity, we drop the momentum-conserving delta function for one-body current operators. With these conventions, the LO axial current takes the following form:
\begin{equation}
\vec{A}_-^{(0)}(\vec{p}\,',\vec{p}\,) = -g_A\sum_i\vec{\sigma}_i\tau_{i,-}\, ,
\label{eqn:LOCurrent}
\end{equation}
where $\vec{\sigma}_i$ is the spin Pauli matrix of nucleon $i$ and $\tau_- \equiv (\tau_x - i\tau_y)/2$ acts on the isospin. 

By NDA, NLO axial currents vanish. At {\NNLO}, there are two types of contributions. One comes from the {\NNLO} correction to the nucleon axial form factor, which is proportional to $\langle r_A^2 \rangle q^2$. With the axial mean-square radius $\langle r_A^2 \rangle \simeq 0.4\, \text{fm}^2$ and the lepton-deuteron momentum transfer $q \sim 1$ MeV, $\langle r_A^2 \rangle q^2 \sim 10^{-5}$; therefore, this part, although nominally {\NNLO}, is negligible. The other part is what we take into account: the $1/m_N^2$ correction to the nucleon axial vector coupling~\cite{Park:1993jf, Long:2010kt, Baroni:2015uza},
\begin{equation}
\vec{A}_-^{(2)}(\vec{p}\,',\vec{p}\,) = \frac{g_A}{2m_N^2}\sum_i \left[\vec{K}^2\vec{\sigma}_i - (\vec{\sigma}_i \cdot \vec{K})\vec{K}\right]\tau_{i,-},
\label{eqn:N2LOCurrent}
\end{equation}
where $\vec{K} = \frac{1}{2}(\vec{p}+\vec{p}\,')$. Here the $\vec{q}-$dependent terms have been neglected due to the smallness of $q$ in near-threshold reactions. An equivalent expression for $\vec{A}_-^{(2)}$ can be found in Ref.~\cite{Krebs:2016rqz}. For expressions of the axial currents in coordinate space, we refer to Refs.~\cite{Park:2002yp, Baroni:2018fdn}.

We find that the following two-body contact axial current operator, as predicted in Ref.~\cite{PavonValderrama:2014zeq}, is enhanced in comparison with NDA:
\begin{equation}
    \vec{A}_{ct}(\vec{p}\,',\vec{p}\,) = \hat{d}_R\,\vec{\sigma}_1\times\vec{\sigma}_2 \left(\pmb{\tau}_1\times\pmb{\tau}_2\right)_-\, .
\label{eqn:Act}
\end{equation}
The LEC $\hat{d}_R$ is usually determined by fitting to observables of a three-nucleon system, e.g., tritium $\beta$ decay~\cite{Park:2002yp} or binding energy~\cite{Marcucci:2013tda}, making use of the relation between $\hat{d}_R$ and the LEC $c_D$ that appears in three-nucleon forces, as demonstrated in Ref.~\cite{Gazit:2008ma}.

The deuteron wave function is yet another essential ingredient. In coordinate space it has the following form:
\begin{equation}
\psi_d^M(\vec{r}\,) = \sum_{L=0, 2}\frac{u_L(r)}{r}\mathcal{Y}_{1L1}^{M}(\hat{r})\zeta_0^0 \, ,
\label{deuteron:function}
\end{equation}
where $\mathcal{Y}_{JLS}^{M}(\hat{r})$ are the normalized spin-angle wave functions~\cite{1979Theoretical2}. The $S$- and $D$-wave components of the wave function $u_0(r)$ and $u_2(r)$ are normalized so that
\begin{equation}
\int_0^\infty dr \left[u_0^2(r)+u_2^2(r)\right] = 1\, .
\end{equation}
We follow Ref.~\cite{Long:2011xw} regarding power counting of the chiral forces in the coupled channel of $\csd$, which actually coincides with NDA up to {\NNLO}. The procedure spelled out in Ref.~\cite{Shi:2022blm} is followed to determine the values taken by the contact LECs in the potentials. We also use the cutoff values adopted in Ref.~\cite{Shi:2022blm}, discarding some cutoff ranges where the numerical accuracy may suffer. 

The matrix element of the axial current between the $pp$ scattering and deuteron states is usually parametrized as
\begin{align}
\left< \psi_d^M\vert A_-^i\vert \psi_{pp} \right> = \delta_{Mi}\sqrt{\frac{32\pi}{\gamma^3}}g_AC_0\Lambda_R(p)\, ,
\label{ME:parametrized}
\end{align}
where $\gamma$ = 45.7 MeV is the deuteron binding momentum, $C_0 = \sqrt{2\pi\eta/(e^{2\pi\eta}-1)}$ the Gamow penetration factor (not to be confused with the contact coupling constants of the chiral potentials), and $\Lambda_R(p)$ the radial matrix element at the $pp$ relative momentum $p$. Expansion of the matrix element consists of contributions from currents and corrections to the wave functions:
\begin{equation}
\begin{split}
    \langle \psi_d^M\vert \vec{A}_-\vert \psi_{pp} \rangle = &\langle {\psi_d^M}^{(0)}\vert \vec{A}_-^{(0)}\vert \psi_{pp}^{(0)} \rangle+ \langle {\psi_d^M}^{(0)}\vert \vec{A}_-^{(0)}\vert \psi_{pp}^{(1)} \rangle\\
    &+ \langle {\psi_d^M}^{(2)}\vert \vec{A}_-^{(0)}\vert \psi_{pp}^{(0)} \rangle+ \langle {\psi_d^M}^{(0)}\vert \vec{A}_-^{(0)}\vert \psi_{pp}^{(2)} \rangle\\
    &+ \langle {\psi_d^M}^{(0)}\vert \vec{A}_-^{(2)}\vert \psi_{pp}^{(0)} \rangle+\cdots \, .
\end{split}
\end{equation}
At LO, the contribution to $\Lambda_R$ from the one-body axial current operator \eqref{eqn:LOCurrent} reduces to the following integral~\cite{Schiavilla:1998je}:
\begin{equation}
    \langle {\psi_d^M}^{(0)}\vert \vec{A}_-^{(0)}\vert \psi_{pp}^{(0)} \rangle = \Lambda_R( p\, \vert \vec{A}^{(0)}_{-} ) =
    \sqrt{\frac{\gamma^3}{2p^2}}\frac{e^{i\delta_{sc}^{(0)}}}{C_0}\int_0^\infty dr\, u_0(r)\chi_0(r;p)\, ,
\label{eqn:LambdaROfA0}
\end{equation}
where $u_0$ and $\chi_0$ are the LO radial wave functions.
At {\NNLO}, the contribution from the axial current~\eqref{eqn:N2LOCurrent} is given by
\begin{equation}
\begin{split}
     \langle {\psi_d^M}^{(0)}\vert \vec{A}_-^{(2)}\vert \psi_{pp}^{(0)} \rangle &= \Lambda_R( p\, \vert \vec{A}^{(2)}_{-} ) =
    \frac{1}{12m_N^2}\sqrt{\frac{\gamma^3}{2p^2}}\frac{e^{i\delta_{sc}^{(0)}}}{C_0} \\
    &\quad \times \int_0^\infty dr \left[ u_0'' \chi_0 + u_0\chi_0'' - 2\left(u_0' - \frac{u_0}{r}\right)\left(\chi_0' - \frac{\chi_0}{r}\right) \right]\, .
\end{split}
\end{equation}

We are left with the matrix elements of the LO current between high-order wave functions: $\langle {\psi_d^M}^{(\nu)}\vert \vec{A}_-^{(0)}\vert \psi_{pp}^{(\nu')} \rangle$, which are in turn induced by higher-order chiral potentials. In much the same way the $pp$ scattering amplitude was expanded [see Eqs.~\eqref{eqn:VstrExpan} and \eqref{eqn:TscExpan}], we can obtain the potential-corrected $\Lambda_R$ through numerical Taylor expansions. First, an auxiliary potential is defined by introducing the dummy parameter $x$:
\begin{equation}
    V(x) = V^{(0)} + x V^{(1)} + x^2 V^{(2)} \, .
\end{equation}
Second, a generating function is calculated through Eq.~\eqref{eqn:LambdaROfA0}, $\Lambda_R(p; x |\vec{A}^{(0)}_-)$. Its Taylor series around $x = 0$ yields desired corrections:
\begin{equation}
\Lambda_R(p; x |\vec{A}^{(0)}_-) = \Lambda_R^{(0)}(p) + x\Lambda_R^{(1)}(p) + x^2\Lambda_R^{\text{pot}}(p) + \cdots\,,
\label{ME:expansion}
\end{equation}
where
\begin{align}
    \Lambda_R^{(1)}(p) &= \langle {\psi_d^M}^{(0)}\vert \vec{A}_-^{(0)}\vert \psi_{pp}^{(1)} \rangle \\
    \Lambda_R^{\text{pot}} &= \langle {\psi_d^M}^{(2)}\vert \vec{A}_-^{(0)}\vert \psi_{pp}^{(0)} \rangle + \langle {\psi_d^M}^{(0)}\vert \vec{A}_-^{(0)}\vert \psi_{pp}^{(2)} \rangle \, .
\end{align}
In practice, construction of the auxiliary potential $V(x)$ can be tweaked if higher numerical accuracy can be achieved or more information is needed. For instance, one can use instead
\begin{equation}
    V(x, y, z) = V^{(0)} + x V^{(1)}_{\cs{1}{0}} + y V^{(2)}_{\cs{1}{0}} + z V^{(2)}_{\csd} \, ,
\end{equation}
which makes the iterative contribution from $V^{(1)}$ and the first-order perturbation of $V^{(2)}$ in two $S$ waves be separated from each other. This breakdown of contributions is unambiguous up to {\NNLO}, where different partial-wave potentials do not mix. We come back to this in Sec.~\ref{sec:results}.

\section{Results and Discussions\label{sec:results}}

Electroweak reactions can reveal rich structure in nuclei. However, multiple low-energy scales often coexist in these reactions, which may call for additional care in EFT analysis. The characteristic scales in $pp$ fusion include the $pp$ initial relative momentum, the deuteron binding momentum $\gamma \simeq 46$ MeV, and the inverse Bohr radius $\alpha m_N \simeq 7$ MeV. At energies of solar-physics interests, $p \lesssim \alpha m_N$, so that the Coulomb potential must be treated nonperturbatively. Therefore, with $\mhi \simeq \delta$ the acceptable upper bound for EFT truncation error at the $\nu$th order will be $(\gamma/\delta)^{\nu + 1}$.

The NDA estimation of the current operators could be upset by enhancement of nonperturbative nuclear dynamics in the initial or final states. We can be alerted to this sort of enhancement by RG analysis as a diagnostic tool. Our strategy of testing NDA of axial current operators against RG invariance is similar to that of Ref.~\cite{Shi:2022blm}. Long-range physics--- contributions from one-body and pion-exchange currents--- are assumed to follow NDA, and we study whether those contributions are independent of the cutoff value $\Lambda$.

\begin{figure}[htbp]
    \centering
    \includegraphics[scale=1.2]{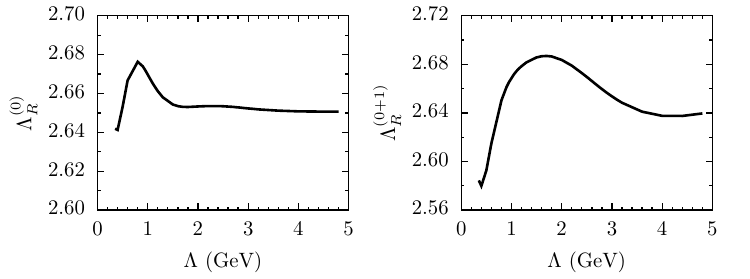}
    \caption{The LO ($\Lambda_R^{(0)}$) and NLO ($\Lambda_R^{(0)} + \Lambda_R^{(1)}$) radial matrix elements for $p=2.17$ MeV as functions of the cutoff value $\Lambda$.
    }
    \label{fig:LONLOcutoff}
\end{figure}

Choosing the initial relative momentum $p = 2.17$ MeV, we illustrate in Fig.~\ref{fig:LONLOcutoff} the cutoff variation of the radial matrix element $\Lambda_R(p)$ at LO and NLO. Cutoff independence is evidently achieved at LO for large cutoff values. The NLO fluctuation appears to be oscillating with a decaying magnitude. The magnitude--- from peak to trough--- is about (2.68 - 2.64)/2.65 $\simeq 1.5\%$, comparable or smaller than the theoretical uncertainty expected of a legitimate NLO $\simeq (\gamma/\delta)^2 \simeq 3\%$. Therefore, we conclude that both LO and NLO are sufficiently insensitive to the cutoff value.

We compare our NLO result with the rates calculated previously in the literature by choosing $p = 0$. With $\Lambda = 1$ GeV and the aforementioned truncation error of $3\%$, the value of the radial matrix element is $2.65 \pm 0.08$. One of the potential model calculation gives $\Lambda^2_R(0)=7.052 \pm 0.007$~\cite{Schiavilla:1998je}, translating to $\Lambda_R(0) = 2.656$. Pionless EFT calculation~\cite{Chen:2012hm} has $\Lambda_R(0) = 2.648$ and NDA-based ChEFT calculation in Ref.~\cite{Acharya:2016kfl} has $\Lambda_R(0) = 2.662$. Our NLO result agrees with these calculations within the uncertainty.

\begin{figure}[htbp]
    \centering
    \includegraphics[scale=1.2]{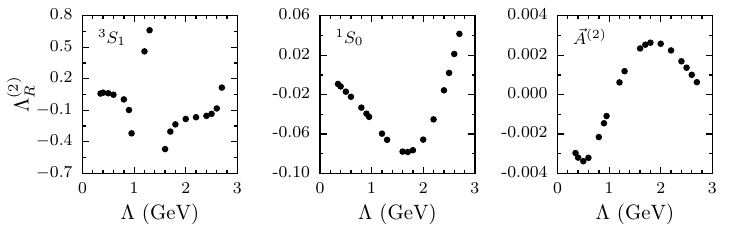}
    \caption{The N$^2$LO corrections of the radial matrix element $\Lambda_R$ as a function of the cutoff $\Lambda$ at $p=2.17$ MeV.
    } 
    \label{fig:N2LOBreakdown}
\end{figure}

At {\NNLO}, the cutoff variation is much more significant. We break down the {\NNLO} corrections at $p = 2.17$ MeV in Fig.~\ref{fig:N2LOBreakdown} according to the source that generates them. ``$\cs{3}{1}$'' is generated by the {\NNLO} deuteron wave function, ``$\cs{1}{0}$'' by the {\NNLO} $pp$ scattering wave function, and ``$\vec{A}^{(2)}$'' by the {\NNLO} axial current operator acting on the LO wave functions. The largest of these variations is due to the {\NNLO} $\csd$ potential, showing as large as $40\%$ deviation with respect to LO based on the values of $\Lambda_R$ for $\Lambda = 1.3$ and 1.6 GeV. The $^1S_0$ potential causes appreciable variation too, with the fluctuation amounting to an uncertainty of $5\%$ based on the values of $\Lambda_R$ for $\Lambda=1.7$ and 2.7 GeV. $\vec{A}^{(2)}$ only probes the cutoff variation of the LO wave functions, which is negligible in comparison with the other two contributions. We notice that both variations of $\cs{3}{1}$ and $\cs{1}{0}$ are much larger than the acceptable {\NNLO} uncertainty $(\gamma/\delta)^3 \simeq 0.4\%$. 

The sensitivity to the cutoff value at {\NNLO} suggests that modification of NDA-based power counting be in order. More specifically, we need to assign the contact axial current $\vec{A}_{ct}$ to {\NNLO} instead of the NDA counting of {\NNNLO}. This is in approximate agreement with the conclusion of Ref.~\cite{PavonValderrama:2014zeq}, where $\vec{A}_{ct}$ was found to be N$^{7/4}$LO based on the analysis using the asymptotic wave functions at short distances.

While Ref.~\cite{PavonValderrama:2014zeq} and our work both agree that RG invariance requires counting of the axial vector current move away from NDA, there are some differences. First, the subleading chiral potentials and current operators are included in the present paper, while Ref.~\cite{PavonValderrama:2014zeq} focused on analyzing the short-range behavior of the LO wave functions.
Second, in our work, however long-range corrections are counted, e.g., LO one-body currents combined with {\NNLO} wave functions, the counterterms that renormalize those long-range matrix elements are assigned the same counting. By this working principle, we will not have fractional indices like in Ref.~\cite{PavonValderrama:2014zeq}.

\begin{figure}
    \centering
    \includegraphics[scale=1.2]{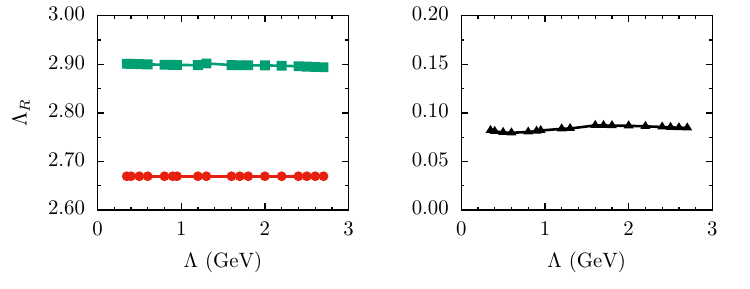}
    \caption{Renormalized {\NNLO} $\Lambda_R$ ($\Lambda_R^{(0)} + \Lambda_R^{(1)} + \Lambda_R^{(2)}$) for various CoM momenta $p$ as a function of the cutoff value. The solids circles, squares, and triangles correspond to $p = $2.17, 10, and 100 MeV, respectively.
} 
\label{fig:RenormalizedNNLO}
\end{figure}

We now demonstrate that $\vec{A}_{ct}$ indeed renormalizes $\Lambda_R$ at {\NNLO}. To determine $\hat{d}_R$, we require the recommended value of $\Lambda_R(p = 0)$ 2.652, provided by Ref.~\cite{Adelberger:2010qa}, to be reproduced at {\NNLO}. Then the prediction of $\Lambda_R$ at other relative momenta is made for various cutoff values. Shown in Fig.~\ref{fig:RenormalizedNNLO}, $\Lambda_R$ is evidently renormalized. 

\section{Summary\label{smry}}

We continue the RG-based analysis of nuclear electroweak currents that was initiated in Ref.~\cite{Shi:2022blm}. Proton-proton fusion is the focus of the present paper. We have calculated the nuclear matrix element of the axial current for this process up to {\NNLO}.

The chiral forces responsible for $pp$ $S$-wave interactions and for the deuteron wave function were constructed according to the power counting laid out in Refs.~\cite{Long:2012ve, Long:2011xw}. Because the incoming $pp$ state is near threshold, the Coulomb force is fully iterated at LO. We have verified numerically that the inclusion of the Coulomb potential does not spoil RG invariance, but the $\cs{1}{0}$ contact terms need to be redetermined by fitting to $pp$ phase shifts.

The novelty of our calculations is perturbative treatment of subleading chiral nuclear forces, as opposed to indiscriminate summation of LO and higher orders. Thanks in large part to strict perturbative calculations, we were able to isolate the contributions from different partial waves and to investigate their cutoff dependence individually. 

At LO and NLO, no significant cutoff variations were found, and our NLO value of the radial matrix element is in agreement with previous calculations within the EFT uncertainty. At {\NNLO}, the chiral force in $\csd$ was found to generate the most cutoff-sensitive contribution. (Interestingly, this is similar to Ref.~\cite{Shi:2022blm} where the $\csd$ potential at {\NNLO} was also found to drive a significant cutoff variation.) As a result, we concluded that one of the two-body contact axial current operators--- defined as $\vec{A}_{ct}$ in Eq.~\eqref{eqn:Act}--- must appear no later than {\NNLO}, one order lower than assessed by NDA. Renormalized by $\vec{A}_{ct}$, $\Lambda_R$ was illustrated to fulfill RG invariance at {\NNLO}. Our finding echos partly the analysis of contact electroweak currents in Ref.~\cite{PavonValderrama:2014zeq}, where $\vec{A}_{ct}$ was assigned N$^{7/4}$LO.

The most immediate consequence of promoting $\vec{A}_{ct}$ concerns the theoretical uncertainty of $pp$ fusion in chiral EFT. Without a reliable input of its LEC $\hat{d}_R$, the $pp$ fusion cross section can be predicted only up to NLO, with an uncertainty conservatively estimated to be $(\gamma/\delta)^2 \simeq 3\%$.

\acknowledgments

We thank Chen Ji for useful discussions. This work was supported by the National Natural Science Foundation of China (NSFC) under Grant Nos. 11735003 and  12275185 (B.L.) and the Fundamental Research Funds for the Central Universities (S.L.).

\bibliography{PPFusionRefs.bib}

\end{document}